\documentstyle[12pt,aasms,epsf]{article}

\begin{document}
\def\half{\textstyle {1 \over 2}   \displaystyle}
\def\fourth{\textstyle {1 \over 4} \displaystyle}

\title{\bf
ON THE 110 keV FEATURE FROM A0535+26:\\
Direct Evidence For A Near-Critical Magnetic Field.} 

\author{
Rafael A. Araya. \\
Dept. of Physics and Astronomy, The Johns Hopkins University,\\
Homewood campus, Baltimore, MD 21218.
\and
Alice K. Harding. \\
Laboratory for High Energy Astrophysics,\\
NASA Goddard Space Fight Center, Greenbelt, MD 20771.
      }
\keywords{line: formation --- magnetic fields --- radiative tranfer ---
	  stars: neutron --- X-rays: stars}

\begin{abstract} 
     A recent high resolution measurement of an absorption line at
 $\omega \simeq$ 110 keV in the phase-averaged spectrum of A0535+26
 (Grove et al. 1994) and the conspicuous absence of a previously reported
 harmonic feature at $\omega \sim$ 50 keV (Kendziorra et al. 1992, 1994) are
 indicative of cyclotron scattering in a magnetic field of about $10^{13}$G.
 However, controversy has risen over whether a lower field scenario may
 account for the alleged absence or weakness of a lower energy fundamental
 harmonic feature.  This work explores these two alternatives through the
 generation of theoretical cyclotron spectra.  For lower field models, 
 a cylindrical geometry of the line forming region and the strong angle
 dependence of the cross section for resonant scattering conspire to
 fill in a first harmonic at $\omega \simeq$ 57 keV. Higher field models
 (B $\simeq$ 10.75 TG), however, yield significantly better fits 
 (${\cal X}^2_{low}/{\cal X}^2_{high} \sim 71$), thus strengthening the case for
 a near-critical field ($B_{crit} \simeq 44$ TG). Phase-resolved OSSE
 spectra are not yet available but would greatly help in resolving this
 issue.
\end{abstract} 
 
\section{INTRODUCTION}

     Notwithstanding the lack of spectral evidence for fields above $\sim$
 5 TG, near-critical magnetic fields of neutron stars are invoked to explain
 some very rapid pulsar spin down rates through magnetic dipole radiation.
 However, recent spectral observations of the transient
 X-ray pulsar A0535+26 by the OSSE instrument on board the Compton Gamma Ray
 Observatory have furnished the most compelling evidence yet for the strongest
 magnetic field directly observed.  Grove et al. (1994) report the 
 detection of a highly significant (F-test $P < 10^{-15}$; Grove, priv. comm.)
 absorption line feature at
 $\omega \simeq$ 110 keV and the conspicuous absence of an obvious harmonic
 feature at a lower energy.  If due to cyclotron resonant scattering at the
 fundamental energy, this phase-averaged spectrum would confirm 
 the existence
 of a magnetic field $B \simeq \frac{1}{4} B_{crit}$ for this object. HEXE 
 observations of an earlier outburst (Kendziorra et al. 1994) showed a 
 feature around 100 keV ($\sim 4.5\sigma$) as well as a weak ($\sim 2\sigma$)
 feature at $\omega \simeq$ 50 keV (the HEXE/TTM instruments on board MIR,
 have better low energy response but coarser high energy resolution than OSSE). 
    
     Cyclotron lines are quasi-harmonic spectral features generated by 
 scattering of resonant photons with the electrons of a magnetized plasma.
 The interaction is resonant because the electron's momentum eigenvalues
 perpendicular to ${\bf B}$, $p_{\perp}/m_e = n~(B/B_{crit})$, are discrete.
 The resonant photon energies, ${\omega}_n/m_e =
 [(1 + 2~n~(B/B_{crit})~{\rm sin}^2\theta)^{1/2} - 1] / {\rm sin}^2 \theta$,
 and the scattering profiles depend on the photon's propagation angle with
 respect to the field $\theta$. 
 Consequently, the line shapes are heavily influenced by the plasma geometry.
 As evidenced above, to properly model a source with a near-critical field,
 $B_{crit} = m^2 c^3/e \hbar$, a relativistic treatment is required
 regardless of the electron temperature.
     
     Theoretical cyclotron spectra have recently 
 been investigated by Wang, Wasserman and Salpeter (WWS 1988), Wang et al. 
 (1989) and by Alexander and M\'esz\'aros (AM 1989).  Our scheme improves on
 previous work by making use of $both$ QED magnetic scattering cross sections
 and relativistic angle redistribution of photons. In addition, the process
 of photon spawning (i.e. degrading of photons to lower harmonic energies
 through electron excitation and subsequent multiple decay) is considered in
 detail and up to four harmonics are included.  For low enough densities
 and/or high magnetic fields (the requirement for the applicability of our
 source code), the photon-electron interactions are predominant and collisional
 terms may be ignored.  Specifically, the radiative cyclotron de-excitation
 rate is given by (Latal 1986)
 $~\nu_r =  \alpha m_e~[c^2/\hbar]~(B/B_{crit})^{\beta}~{\rm with}
  ~\beta \in$ [2, .5] for the undercritical and overcritical field limits
 respectively. For $B_{12} \simeq 1$, this becomes
 $~\nu_r \sim  2 \times 10^{15} B_{12}^2 ~{\rm sec}^{-1}$.
 Comparing the latter expression with 
 that from the collisional rate (Bonazzola, Heyvaerts and Puget 1979),
 $\nu_c =  5 \times 10^{8}~(n_e/10^{21}\,{\rm cm^{-3}})\,B_{12}^{-3/2}$,
 and considering that the continuum optical depths, 
 $\tau_c \stackrel {\sim}{<} 5.\times 10^{-4}$,
 in the present work give low column densities:
 $n_l \stackrel {\sim}{<} 7.5\times 10^{20} {\rm cm}^{-2}$,
 the neglect of collisions is readily justified.  
 Moreover, the predominance of the magnetized vacuum motivates one chief
 assumption: the consideration of unpolarized photons only. 
 WWS88 report that averaged (vacuum mode) polarized spectra closely
 resemble the unpolarized case for optical depths of a few.
 
    A0535+26 is an X-Ray binary transient pulsar with a pulse period of 
 104 sec. The orbital period is 111 days and the companion of the neutron
 star is a 12 $M_\odot$ Be star. X-Ray episodes for the system are
 classified into: giant, normal and no burst (Motch et al. 1991). When
 an outburst does occur, it is typically associated with periastron passage
 (Hutchings 1984). Both detections of cyclotron line features were made 
 during giant outbursts.
 Cyclotron line features above 100 keV in the spectra of X-ray pulsars 
 have been difficult to detect for several reasons.  First, high resolution
 detectors with good efficiency at high energies have only recently begun to 
 observe these sources.  Second, the intrinsic spectra of many X-ray
 pulsars turnover at energies above $\sim 20$ keV.
 These facts suggest that even if such
 high fields are common, few may be discovered through their spectral 
 signature.  The hardness of the continuum spectrum of A0535+26 below the
 turnover and its brightness during outbursts makes it an
 unusually favorable source for detection of high-energy line features by 
 the OSSE instrument.

    Hypothetically, two scenarios for the absence or weakness
 of a lower energy feature can be envisioned. In the first case,
 $B \simeq 5$ TG and the geometry of the line forming region combined with
 the strong angle dependence of the scattering cross section conspire
 either to fill up a first harmonic at large angles or to form a
 very shallow and broad fundamental feature at smaller angles.
 In the other case,
 the 110 keV absorption line as a fundamental provides direct evidence for
 a magnetic field $B \simeq 11$ TG.
 In this $Letter$, we explore these two alternatives by modeling the resonant
 transfer of unpolarized photons with a Monte Carlo scheme. 

\section{The February, 1994 event: Giant outburst.}

    A complete account of the February 1994 event can be found in
 Grove et al. (1994).
 The continuum spectrum is modeled as a power law times an exponential:
 $ dN_{\omega}/(d{\omega}d{\mu}) \sim 
   {\omega}^{-\alpha} {\rm exp} (-{\omega}/kT_{c})$. Here $\alpha$ is (minus)
 the power law index, $T_{c}$ is the continuum temperature, and
 $\mu \equiv {\rm cos}\theta$ is the cosine of the viewing angle to the field.
 Their best fit parameters are:$~T_{c} = 17.8~{\rm keV}~{\rm and} 
 ~ \alpha = -.13$. The `second harmonic' (see below) line parameters are:
 $\omega$ = 110 keV, $\tau$ = 1.8 (optical depth), and H.W.H.M. = 24 keV.
 The fits of Grove et al. are consistent with a feature 
 at $\omega \simeq$~55 keV 
 having  $\tau \simeq$~.1 and H.W.H.M. $\simeq$ 14.2 keV. Nevertheless, as 
 pointed out in their paper, the existence of such a lower energy feature 
 is not strictly required for reasonable ${\cal X}^2$ fits to the data. It 
 serves mostly to flatten the spectrum at low energies.

\section{A brief description of the code.}

     Our code makes use of the cyclotron scattering cross section
 calculated as a 2nd order QED process. The cross section exhibits strong
 angular dependence and 2nd order effects allow off mass-shell
 states to contribute to the scattering
 (Harding and Daugherty 1991; hereafter HD91).

    Natural line widths of the cross sections are introduced by using
 dressed electron propagators and a choice of eigenbasis and spin states
 which diagonalizes the electron's self-interaction. Thus, cartesian
 coordinate solutions to the Dirac-Landau equation which are eigenstates
 of the Sokolov-Ternov spin operator (Herold, Ruder and Wunner 1982)
 and which diagonalize the mass operator are used.  The dressed propagator
 (i.e. inverse of the D-L equation with a self-energy term) introduces small
 imaginary (and real) corrections to the resonant energies (Graziani,
 Harding and Sina 1995). To first order in $\alpha$,
 the imaginary part of the correction amounts to the substitution:
 $\omega \rightarrow \omega - i \frac{\Gamma}{2}$ everywhere in the S matrix.
 Here $\Gamma$ is the spin dependent relativistic cyclotron transition rate.
 Upon squaring of the S matrix element a quasi-Lorentzian line width results 
 (HD91 eq. 20).
 This prescription is used in our code for the scattering mean-free paths 
 while a Lorentzian line width approximation (absorption approximation,
 following HD91)
 allows a semi-analytical inversion of the partial profile convolution 
 integrals to sample the scattering electron momentum $P_{z}$.

    Furthermore, the code uses a relativistic Maxwellian electron distribution
 function and includes excitation of Landau levels up to $n = 4$ and their
 subsequent de-excitation and photon spawning.  An approximate relativistic
 scattering angle redistribution of photons allows for mixing between angle
 bins.

\section{Model Calculation}

    Two geometries for the scattering region are considered: plane slab (with
 the magnetic field parallel to the slab normal) and cylindrical (with the 
 field parallel to the cylinder axis). An isotropic
 continuum photon spectrum is incident from a source at the slab midplane
 or along the cylinder axis.  The escaping photons are accumulated in
 four ranges of $\mu$ (the cosine of the viewing angle to the field).
 Due to the steepness of the input spectrum, the photons are injected
 with a flat spectral distribution and are assigned appropriate weight
 factors. This improves the statistics, but upscattered photons with 
 large weight factors create some noise level in the high energy part 
 of the spectrum.  Following Grove et al. (1994),
 there are two parameters to describe the continuum: $\alpha$ and $T_{c}$.
 Our Monte Carlo code takes three parameters to characterize the line:
 the electron temperature $T_e$, the local 
 magnetic field $B$ (assumed uniform), and the minimum value of continuum
 optical depth $\tau_{c}$ (in the direction parallel to $B$ for a slab
 and perpendicular to $B$ for a cylinder). A summary of all parameter
 combination trials is given in Table 1. A representative sample of runs
 as well as more detail analysis of the trial runs leading to our best fits 
 was presented in Araya and Harding (1996), hereafter AH96-I. 

     The differences in the line features with various combinations of
 parameters result from an interplay between the geometry and the
 variation in the cross section with the photon's angle of propagation
 with respect to the field.  The line profiles are narrower and deeper
 at large angles, due in part to one dimensional Doppler broadening
 (cf. HD91). Concurrently, for outgoing photons
 in a cylindrical scattering region the optical depth at small angles
 is largest [$\tau^{cy}_c(\theta) = \tau_c / {\rm sin}\theta$]  
 while for photons emerging from a slab the optical depth is largest
 at large angles [$\tau^{sl}_c(\theta) = \tau_c / {\rm cos}\theta$]  
 Thus, escaping photons in cylindrical scattering regions tend to
 redistribute in the line wings from small to large angles, where absorption
 features may be filled in, while filling in by angle redistribution is much
 less effective in slab geometries.  The (unnormalized) value of $\tau$ at the
 line center may be readily evaluated from the continuum depth:
 $\tau(\theta) = \tau_c(\theta) * \sigma_{scatt}^{tot} / \sigma_{Thomson}$.

    Standard estimates for $T_e$ from the doppler line width,
 $\Delta\omega_{dop} = \omega_{n} (2kT_e/mc^2)^{1/2}{\rm cos}\theta$,
 are inadequate for
 two reasons: first, relativistic kinematics produces broadening even at
 $\theta = 90^0$; second, fundamental features have complex structure 
 resulting from angle mixing and extra injection from spawned photons.
 Detailed analysis of thermal balance in the line forming region has
 been addressed by Lamb, Wang and Wasserman (1990), hereafter LWW90,
 by assuming classical magnetic resonant and non-resonant Compton cross
 sections.  Following the single scattering formalism, assuming line
 dominated cooling and heating, and assuming isotropic injection, an
 estimate for the electron temperature yields (eq. 50 of LWW90):
 $T_e = \omega_{\rm cyc} / (2+\alpha_{eff})$, with $\alpha_{eff}$
 being (minus) the effective power law index at the line. For 
 $\tau \gg 1$, LWW90 produce a numerical estimate from Monte Carlo runs: 
 $T_e \sim \fourth\omega_{\rm cyc}$ for $\alpha_{eff} \sim 1$. The latter
 estimate is not justified for A0535+26 where the continuum at the
 line(s) has a much larger effective index, 
 $\alpha_{eff} \sim 6.5$ (3.6) at $\omega = 110$ keV (55 keV), and 
 where the optical depth is close to unity.
 Although we observed that the relativistic and non-relativistic 
 resonant cross sections for $B = 0.24~B_{crit}$ 
 differ by as much as 50\% one line width from resonance, the above 
 estimate is used as a guide: $T_e \simeq$ 12.9 keV and 9.8 keV for the
 high and low field models respectively.  Nevertheless,
 in the present work $T_{e}$ is considered a free parameter determined
 empirically from the emerging line features.
 To this end, account is made for the hardness of the incoming continuum
 as well as the emergent flux distribution   
 in the wings of the fundamental harmonic.
 Further discussion on the evaluation of $T_e$ is given in AH96-I.

\section{Results} 

\subsection{The lower field case} 

     Two foremost restrictions on the emergent theoretical spectra 
 (illustrated in fig 1a and 1c) must be
 imposed to comply with the data: the first harmonic line must have 
 E.W. $\ll$ 1, and the second harmonic feature must be very wide and
 much more significant ($\tau \sim$ few).
 
     Since the scattering line profile is larger at the first harmonic than
 at the second, some process must work to fill in the fundamental line
 feature. Photon spawning and angle redistribution of resonant photons may
 both enable partial filling of a fundamental feature. Nevertheless, the soft
 continuum from this source inhibits photon spawning since the number of
 second harmonic photons is only a small fraction ($\stackrel{\sim}{<} 1\%$)
 of the first harmonic photons (the spawned photon contribution is shown in
 the lower portion fig. 2).  On the other hand, angle redistribution from
 small to large angles does permit filling in of the fundamental absorption 
 feature for cylindrical geometries since the profiles are narrower there
 (lower portion of fig 1c). In this case, however, overfilling occurs
 for optical depths compatible with the second harmonic feature, thus
 producing a sharp $emission$ feature at 59 keV as well as a second harmonic
 line that is too narrow (and very weakly dependent on higher values
 of $T_e$).

    Alternatively, spectra for $\mu \sim [.5-.75]$ in both geometries
 (fig 1a and 1c) exhibit a fairly shallow fundamental feature and a broader
 second harmonic. For this angle bin, slab geometries display
 slightly stronger emission wings at the fundamental and a deeper
 second harmonic with a sharper transition into the high energy 
 continuum compared to their cylindrical counterparts. 
 Nonetheless, the width
 of the feature at $\omega \sim$ 110 keV still fails to match the 
 observation for electron temperatures consistent with spectral
 hardness, and a significant feature prevails at the fundamental energy.
 Furthermore, significantly increasing the electron temperature: 
 $T_e \gg  10$ keV (the value implied by eq. 50 of LWW90) enables
 the first harmonic to become                   
 more prominent and fails to considerably broaden the second harmonic.
 (see AH96-I).  Lastly, we note that increasing $\tau$ deepens the
 fundamental, while decreasing it obliterates the feature at 110 keV.

    While our model is able to roughly produce either an `undetectable'
 first harmonic line or a broad second harmonic, the parameter space for
 the $mutual$ satisfaction of these two conditions is virtually null 
 (we qualify this statement in the conclusions).
 The best fit lower field model, shown in figure 2, is therefore a
 compromise between the two constraints mentioned above. 

\subsection{The higher field case}

     Figures 1b and 1d illustrate that the closest line feature to that of
 A0535+26 is provided by a cylindrical geometry in angle bin $\mu > .75$. 
 Our best fit, shown in figure 2 along with the OSSE data, 
 is obtained under these conditions (see caption for details). Note that
 the value $T_e$ = 12.94 obtained from the single scattering estimate
 of LWW90 is in rough agreement with the fitted value of 12 keV. Moreover, 
 the magnetic field strength $B = 10.75$ TG exceeds the naive estimate based
 on placing the line centroid at 110 keV: $B \in [9.5,~10.5]$ TG for
 $\mu \in [0,~1]$. 

    Several remarks should be made regarding the quality of the fit.
 The spectrum from the OSSE instrument has very small systematic
 uncertainties (5\% of the flux for $\omega < 80$ keV and 3\% for  
 $\omega < 110$ keV (Grove, 1995, priv. comm.)) and our Monte Carlo 
 spectra have statistical noise fluctuations that easily exceed those
 in the data. The only apparent mismatch for the high field model 
 occurs for the gentler slope of the observed line feature for energies below
 110 keV when compared with the theoretical spectrum. One must view this
 slight discrepancy in light of the limitations of the observation 
 (the data is a phase averaged spectrum) and the limitations of the code
 (e.g. resulting spectra which are averages over angle bins, representative
 geometries, etc).
 We must emphasize that our best fits have been made using the published
 flux readings from the OSSE instrument as `rigid' data points, and $not$ 
 through the more formal, forward-folding deconvolution of the model with the 
 instrument response function (A scheme for less computationally exhaustive 
 fitting of observational data is not yet available).
 Therefore, both estimates for the goodness of fit are to be taken not as
 absolute formal numbers but rather as relative values.
 The value of the reduced ${\cal X}^2$ is 8.1 for the best high field model
 and 573 for our best low field spectrum (both shown in fig 2).
 
\section{Conclusions}

    The present work indicates a very small likelihood that a lower field
 model would acceptably fit the spectrum of A0535+26.  The observational
 evidence does not seem to support the weak, albeit seemingly detectable,
 fundamental line that would result at energies of $\sim$ 58 keV. However,
 the OSSE instrument's low energy threshold of 45 keV hampers a definite 
 assessment of a line `non-detection'.  The TTM/HEXE instruments
 may be better equipped for this lower energy measurement.  
 Exactly what is meant by an `un-detectable' line feature by a particular 
 X-ray instrument is not a trivial question.  Work in progress using the
 HEXE phase $resolved$ data has shown the possibility that a 
 fundamental feature with {\em emission wings} may be completely concealed
 by the instrument's response function (Kretschmar et al. 1996).

     Note that the uncertainty introduced by using the phase averaged 
 spectrum is determined by the relative dominance of the highest flux phase,
 if the line is deep enough.  Even though no phase resolved spectra are 
 available from OSSE, work in progress with the HEXE data indicates
 excellent agreement of a formal fit
 (${\cal X}^2_{red} = 1.9$) with an essentially identical
 high field model for the highest flux phase (d) of Kendziorra et al. (1994)
 (P. Kretschmar, priv. comm.).
 This motivates the use of the phase averaged spectrum from OSSE.

    Lastly, since the model's spectral signatures have strong dependencies
 on geometry and on viewing angle with respect to the field, we cannot
 rule out that a very radical geometry or an anisotropic incident photon
 distribution, which we plan to study, may fill up the first harmonic better
 than can be seen from our representative geometries and isotropic injection
 models.

    In this $Letter$ we provide a reasonable appraisal for the various physical 
 parameters and conclude that the 110 keV feature from A0535+26 indeed may
 come from the highest magnetic field known by $direct$ evidence. 
 Further work on phase resolved data is in progress.

   $Acknowledgments$: We would like to thank Ramin Sina for supplying
 the code to calculate the scattering cross section using Sokolov-Ternov
 eigenfunctions, P. Kretschmar and J. E. Grove for valuable discussions on 
 the experimental data and A. Szalay for allowing us access to the 
 supercomputer resources at his disposal at the Johns Hopkins University. 

\newpage

\vspace{0.3cm}
\begin{tabular}{|l||l|l|l|l|l|}
	\hline\hline
	 $T_{e}$ & \multicolumn{5}{c|}{Magnetic Field strength B (TG),
	  ~~ {[$B/B_{crit}$]} }  \\ 
	 (keV) & B = 5.25 & B = 9.5 & B = 10.36 & B = 10.75 & B = 11.14 \\
	  &  [.12] & [.21] & [.23] & [.24] & [.25] \\
        \hline \hline 

 5. & $3._{-3}^{cy}$ & $1._{-3}^{cy,sl}$ & & & \\
  \hline    
 10. & & & & & $5._{-4}^{cy,sl}$ \\     
  \hline    
 12 & $2._{-4}^{cy,sl}$ & & $5._{-4}^{cy,sl}$
   & $5._{-4}^{cy,sl}$ & $5._{-4}^{cy,sl}$ \\     
  \hline    
 12.5 & $3.5_{-4}^{cy}$ & & & & $5._{-4}^{cy}$ \\     
  \hline    
 14 & & & & & $8._{-4}^{cy}~~1._{-3}^{sl}$ \\     
  \hline    
 16 & $2.5_{-4}^{cy}$ & & & &  \\     
  \hline    
 20 & $2._{-4}^{cy,sl}$ & & & & $2._{-4}^{cy,sl}$ \\     
  \hline    
 27 & $1._{-4}^{cy}$ & $1._{-3}^{sl}$ & & &  \\     
	\hline\hline
\end{tabular}

\begin{table}[h,b,t]
	\caption{
	Summary of Parameter Combination Trials. The values of the
	continuum optical depth $\tau_{c}$ and of the geometry are
	enclosed.  Subscripts are the powers of ten for the value
	of $\tau$ (i.e. $4_{-5} = 4 \times 10^{-5}$.)~~ 
	Cylinder = $cy$ and Slab = $sl$ in the superscripts.
	}
\end{table}

\newpage
\begin{figure}
   {\footnotesize
   Figure 1: Sample angle-dependent photon count spectra.
   Each run results from 50000 to 100000 photons injected isotropically.
   $\mu = {\rm cos}\theta$. $Dashed~line$: injected continuum.
   $Solid~line$: output scattered spectrum. $Crosses$: single upscattered
   photons. Points with solid errors are the OSSE phase averaged spectrum.
   Spectra from slabs are on the top and from cylinders on the 
   bottom. Columns share the values of $\tau,~T_e~{\rm and}~B$.
   $T_c=16.8$ keV and $\alpha = 0$ for a and c (lower field runs). 
   $T_c=14.5$ keV and $\alpha = 0$ for b and d (higher field runs). 
   See text for notation. The flux normalization is arbitrary.}
\end{figure}

\begin{figure}
   {\footnotesize
    Figure 2: Best fit theoretical spectra for high field (two top 
    graphs) and low field models (two bottom graphs). The OSSE
    phase-averaged photon count spectrum for A0535+26 is shown as
    points with solid error bars (absolute flux for top
    plot on the right scale).  The $dashed~horizontal~bars$ are the 
    model spectra binned
       according to the energy intervals of the OSSE instrument. 
    The $solid~lines$ are the same modeled spectra but with equally spaced
       energy bins.  The $dashed~line$ is the continuum input spectrum.
  The continuum parameters are: $T_{c} = 19.$ and $\alpha = 0.66$.
  For $B$ = 10.75 TG the line parameters are: 
  angle bin $\mu > .75$, $\tau_{c} = 5 \times 10^{-4}$ ($\tau \simeq 
  3.15$) and $T_{e} = 12$ keV.
  For $B$ = 5.25 TG the line parameters are: 
  angle bin $.75 > \mu > .5$, $\tau_{c} = 2 \times 10^{-4}$ ($\tau \simeq 
  1.26$) and $T_{e} = 12$ keV.
    }
\end{figure}

\pagebreak

\begin{figure}
\centerline{
\epsfbox{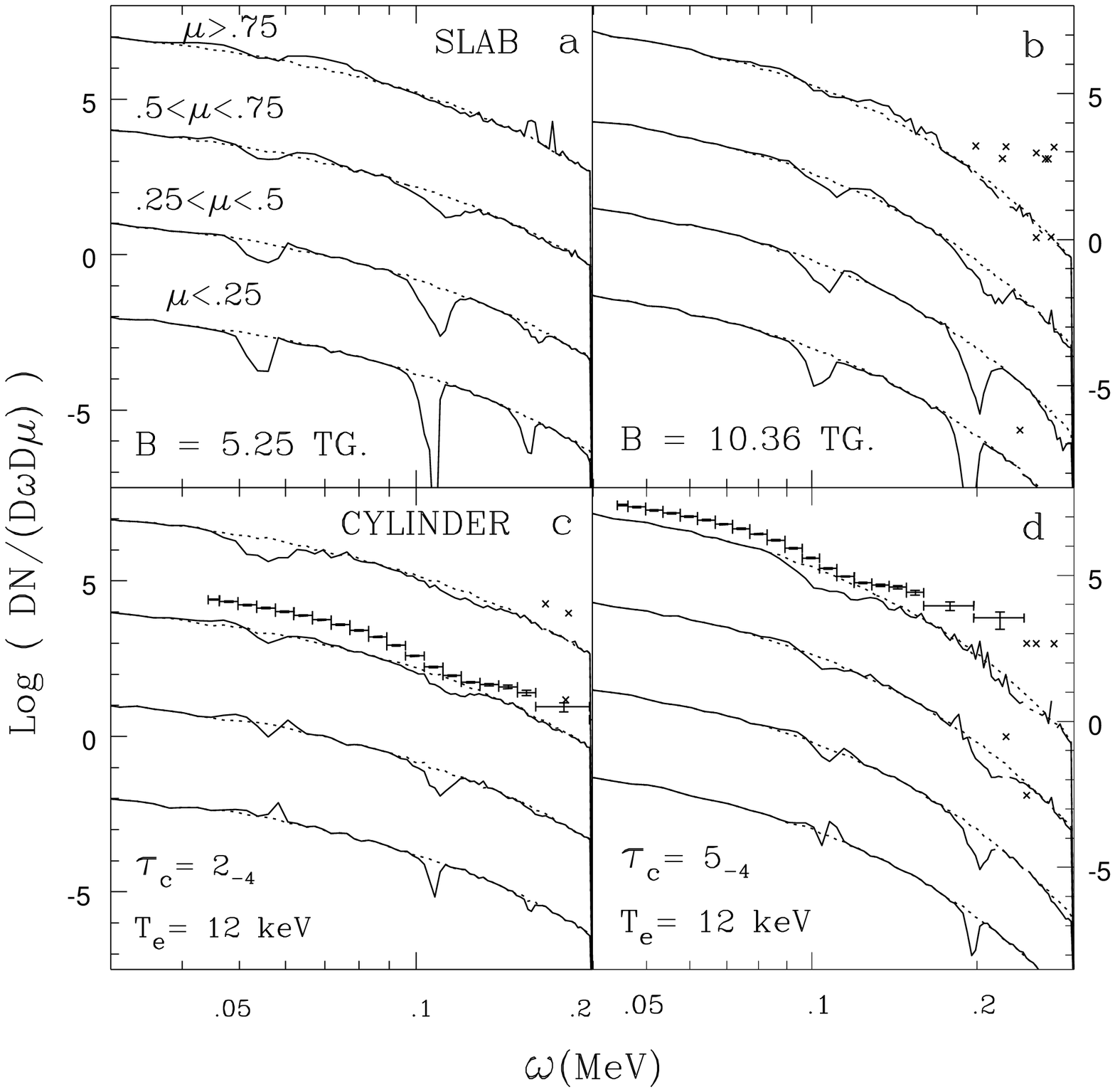}
}
\caption{}
\end{figure}

\newpage

\begin{figure}
\centerline{
\epsfbox{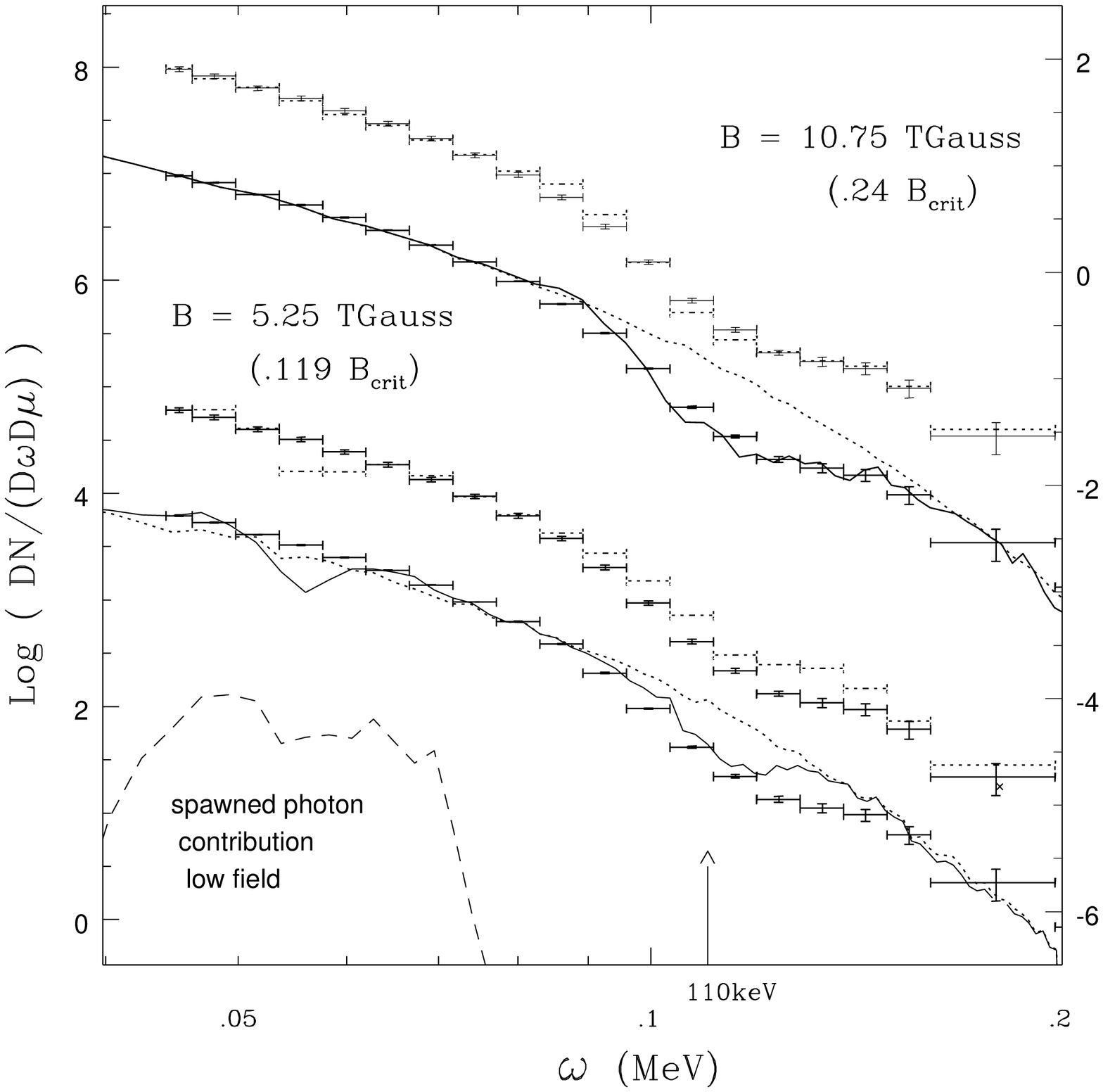}
}
\caption{}
\end{figure}

\end{document}